\documentclass[sn-nature]{sn-jnl}

\usepackage{graphicx}%
\usepackage{multirow}%
\usepackage{amsmath,amssymb,amsfonts}%
\usepackage{amsthm}%
\usepackage{mathrsfs}%
\usepackage[title]{appendix}%
\usepackage{xcolor}%
\usepackage{textcomp}%
\usepackage{manyfoot}%
\usepackage{booktabs}%
\usepackage{algorithm}%
\usepackage{algorithmicx}%
\usepackage{algpseudocode}%
\usepackage{listings}%

\theoremstyle{thmstyleone}%
\theoremstyle{thmstyletwo}%

\theoremstyle{thmstylethree}%

\begin{document}

\title[Article Title]{Observational evidence for Early Dark Energy as a unified explanation for Cosmic Birefringence and the Hubble tension}


\author*[1,2]{\fnm{Joby} \sur{Kochappan}}\email{jobypk@gmail.com}

\author[2,3]{\fnm{Lu} \sur{Yin}}\email{yinlu@shu.edu.cn}

\author[4,5]{\fnm{Bum-Hoon} \sur{Lee}}\email{bhl@sogang.ac.kr}

\author[6]{\fnm{Tuhin} \sur{Ghosh}}\email{tghosh@niser.ac.in}

\affil*[1]{\orgdiv{Manipal Centre for Natural Sciences}, \orgname{Manipal Academy of Higher Education}, \orgaddress{\street{Manipal}, \postcode{576104}, \country{India}}}

\affil[2]{\orgdiv{Asia Pacific Center for Theoretical Physics}, \orgaddress{\city{Pohang}, \postcode{37673}, \country{Korea}}}

\affil[3]{\orgdiv{Department of Physics}, \orgname{Shanghai University}, \orgaddress{\city{Shanghai}, \postcode{200444}, \country{China}}}

\affil[4]{\orgdiv{Center for Quantum Spacetime}, \orgname{Sogang University}, \orgaddress{\city{Seoul}, \postcode{121-742}, \country{South Korea}}}

\affil[5]{\orgdiv{Department of Physics}, \orgname{Sogang University}, \orgaddress{\city{Seoul}, \postcode{121-742}, \country{South Korea}}}

\affil[6]{\orgdiv{National Institute of Science Education and Research}, \orgname{An OCC of Homi Bhabha National Institute}, \orgaddress{\city{Bhubaneswar}, \postcode{752050}, \state{Odisha}, \country{India}}}


\abstract{We test the $n$=3 Ultralight Axion-like model of Early Dark Energy (EDE) with the observations of the $EB$ mode of the cosmic microwave background (CMB) radiation, and local expansion rate measurements. Our results show that the shape of the CMB $EB$ angular power spectrum is sensitive to the background cosmological parameters. We run Markov chain Monte Carlo (MCMC) simulations to fit the $\Lambda$CDM + EDE parameters simultaneously and find that the EDE  model with $n$=3 can provide a good fit to the observed CMB $EB$ spectra,  consistent with the locally measured value of the Hubble constant. Our result is the first to show that axion-like EDE can provide a unified explanation for the observed cosmic birefringence and the Hubble tension.}




\maketitle

\section{Introduction}
\label{sec:intro}

The $\Lambda$CDM model has received strong observational support in the last few decades, but it still faces a few noteworthy challenges. Two of the big challenges to the $\Lambda$CDM model are the Hubble tension and the observation of cosmic birefringence. The Hubble tension is the $\approx 5\sigma$ discrepancy \cite{Riess:2022} found between the value of the Hubble constant today, $H_0$, as inferred from early universe measurements such as the Cosmic Microwave Background (CMB) radiation, and direct measurements from the local distance ladder. Presently, there are numerous efforts to resolve the Hubble tension with new models, as well as in reducing the systematics in the data \cite{Knox:2020,Poulin:2019,Jedamzik:2020,Mazurenko:2024,Lombriser:2020,Jones:2018,Shanks:2019,Cerneiro:2022,Rezazadeh:2022}. However, there is still no widely accepted satisfactory resolution of this conflict. Cosmic birefringence is the rotation of the plane of linear polarization of the CMB photons during their travel from the last scattering surface to the observer \cite{Komatsu:2022}. This effect has received a lot of academic attention \cite{Namikawa:2023,Ferreira:2024,Greco:2024,Namikawa:2024,Kin-Wang:2006}. Recently, this rotation angle was measured, $\beta \approx 0.34^{\circ}$, using the Planck 2018 polarization maps with a statistical significance of 3.6$\sigma$ \cite{Minami:2020,Eskilt:2022}, and provides exciting prospects for new physics beyond the standard $\Lambda$CDM model by considering Early Dark Energy (EDE) \cite{Lue:1999}.

EDE models attempt to resolve the Hubble tension by adding an extra dark-energy-like component which modifies the expansion history at high redshifts, but decays quickly post recombination so as to not affect the late Universe \cite{Smith:2020}. This leads to a larger value of $H_0$ inferred from the CMB data and reduces the tension between the early time and late time measurements, without introducing new tensions in the other parameters. Additionally, in some of the axion-like EDE models, cosmic birefringence is a natural byproduct of the Chern-Simons coupling between the EDE field and the CMB photons. Thus EDE provides an exciting possibility of explaining both the Hubble tension and cosmic birefringence, hitting two birds with one stone.

The EDE component is commonly {modeled} by a pseudo-scalar field such as an ultralight axion (ULA) with a potential $V(\phi) {\propto} \left(1-\cos\left[\phi/f\right]\right)^n$ \cite{Smith:2020}, where $\phi$ is the pseudo-scalar field, $f$ is the axion decay constant and $n>1$ is a phenomenological parameter. There are also scalar field models with $\alpha$-attractor \cite{Braglia:2020} and Rock {`n'} Roll potentials \cite{Agrawal:2023} which were considered in \cite{Yin:2023}. Despite the promising aspects of EDE, recent articles \cite{Eskilt:2023} find that EDE, specifically the ultralight axion-like models with $n=3$, are not supported by the CMB observations. The authors in \cite{Eskilt:2023} fit the $n=3$ model to the Planck 2018 polarization data \cite{Eskilt:2022}, varying the Chern-Simons coupling constant, $g_{EDE}$, and fixing all the other EDE parameters, and found that the shape of the resulting $EB$ power spectra does not agree with the data. This implores us to investigate the effect of varying the EDE parameters instead of fixing them, on the process of fitting the model to the data, which is a key result of this work.

This article is a follow up to our previous paper \cite{Yin:2023} where we compared the $\alpha$-attractor and Rock {`n'} Roll models with the CMB data. In this work, we focus on the pseudo-scalar field models based on ultralight axions with $n=3$. We study the dependence of the CMB $EB$ spectra on the EDE parameters, energy density $f_{EDE}$, critical redshift $z_c$, coupling constant $g_{EDE}$ and initial value $\theta_i$, and simultaneously fit the $EDE+\Lambda$CDM model parameters to the CMB, BAO and $H_0$ data.

This paper is organised as follows. In section \ref{sec:models}, we give a brief review of the axion-like EDE model that we have considered. In section \ref{sec:fit_gede}, we describe the datasets that we have used for this study and fit the coupling constant, $g_{EDE}$, in the axion-like EDE model with $n$=3 to the observed CMB $EB$ power spectrum. Meanwhile, we also show a comparison of our results with the previous results in the literature. We discuss the dependence of the shape of the CMB $EB$ power spectra on the EDE and $\Lambda$CDM model parameters and their implications for the parameter estimation from the data in section \ref{sec:eb_param_dep}. We present our results and best-fit parameter estimates from a full MCMC analysis of the data in section \ref{sec:fit_ede+lcdm}. Finally, in section \ref{sec:7} we draw conclusions from our results and make comparisons with other articles in the literature, and explore directions for future research in the topic.

\section{The axion-like Early Dark Energy models}
\label{sec:models}
We consider axion-like pseudo-scalar field EDE models with potentials given by,

\begin{equation}
    V(\phi) \propto \left(1-\cos\left[\phi/f\right]\right)^n,
\label{eqn:model_gen}
\end{equation}
where, $V(\phi)$ is the field potential, $\phi$ is the axion-like pseudo-scalar field, $f$ is the axion decay constant, and the index $n$ is a phenomenological parameter that takes values $n>1$ to ensure that the EDE density dissipates rapidly after recombination. In the following sections, we will use $\theta = \phi/f$ to simplify the mathematical expressions. These models are motivated by string theory \cite{Witten:2006,Svrcek:2006,Arvanitaki:2010,Marsh:2011,Kamionkowski:2014}, and have been the investigated by numerous articles in the literature \cite{Smith:2020,Yin:2023,Eskilt:2023,Karwal:2016,Poulin:2019,Kable:2024,Hagimoto:2023}. Additionally, they contain a Chern-Simons term coupling the axion field to the CMB photons, which generates a signal of cosmic birefringence in the CMB, in the form of a non-zero correlation between the $E$ and $B$ modes of CMB polarization. This additional coupling term in the Lagrangian density is written as, $-\frac{1}{4}g\phi F_{\mu \nu}\Tilde{F}^{\mu \nu}$. $g$ is the coupling constant which we will refer to as $g_{EDE}$ henceforth, $\phi$ is the pseduo-scalar field, $F_{\mu \nu}$ is the electromagnetic tensor for the CMB photons and $\Tilde{F}^{\mu \nu}$ is its dual. The EDE field behaves like a cosmological constant before a critical redshift $z_c$ and then decays rapidly without making changes to the late time evolution of the Universe.

As explained in literature\cite{Eskilt:2023}, the coupling term induces a difference between the phase velocities of the left and right hand circularly polarised waves, leading to a rotation, $\beta$, of the plane of polarisation depending on the value of the field, $\phi$. As the field, $\phi$, evolves with time, it changes the rotation angle. The change in the rotation angle $\beta$ from time $t_1$ to time $t_2$ is given by \cite{Eskilt:2023},

\begin{equation}
    \beta (t_1, t_2) = \frac{g_{EDE}}{2}(\phi(t_2) - \phi(t_1))
    \label{eqn:beta}
\end{equation}

The CMB polarisation power spectra can then be written as \cite{Murai:2023},

\begin{equation}
    \mathcal{C}_\ell^{XY} = 4\pi \int \textrm{d(ln q)}\mathcal{P}_s(\textrm{q})\Delta_{X,\ell}(\textrm{q})\Delta_{Y,\ell}(\textrm{q}),
\end{equation}
where, $\mathcal{P}_s$ denotes the primordial scalar perturbations power spectrum, $X$ and $Y$ are labels for the $E$-mode and $B$-mode of the CMB polarisation, and $\Delta$s are the Fourier transforms of the Stokes parameters of linear polarisation as explained in \cite{Murai:2023}. Under the simplifying assumption of a constant $\beta$, the CMB polarisation power spectra can be reduced to,

\begin{equation}
    \mathcal{C}_\ell^{EE} = \cos^2(2\beta)\Tilde{\mathcal{C}}_\ell^{EE} + \sin^2(2\beta)\Tilde{\mathcal{C}}_\ell^{BB},
\end{equation}

\begin{equation}
    \mathcal{C}_\ell^{BB} = \cos^2(2\beta)\Tilde{\mathcal{C}}_\ell^{BB} + \sin^2(2\beta)\Tilde{\mathcal{C}}_\ell^{EE},
\end{equation}

\begin{equation}
    \mathcal{C}_\ell^{EB} = \frac{1}{2}\sin(4\beta)\left(\Tilde{\mathcal{C}}_\ell^{EE} - \Tilde{\mathcal{C}}_\ell^{BB}\right),
\end{equation}
where, $\Tilde{\mathcal{C}}_{\ell}$ denote the power spectra for the case, $g_{EDE}=0$.
In this article, we focus on the axion-like potentials given by equation \ref{eqn:model_gen} with $n$=3.

\section{Fitting $g_{EDE}$ with the CMB $EB$ spectrum}
\label{sec:fit_gede}
In this section, we fit the parameter, $g_{EDE}$, with the observed CMB $EB$ angular power spectrum. As demonstrated in \cite{Naokawa:2023}, it is important to consider the effect of gravitational lensing on the CMB polarization spectra induced by axion-like particles, so we use the lensed CMB spectra in our analysis. We use the stacked $EB$ power spectrum given in \cite{Eskilt:2022} as the observations. We fit the model described in equation \ref{eqn:model_gen} with $n$=3, and follow the same binning strategy as \cite{Eskilt:2022} to bin the multipoles from $\ell$=51 to 1490 in equal bins of width $\Delta \ell$=20. We fix all other parameters to the best-fit values found in \cite{Poulin:2019} and mention them in Table \ref{tab:bestfit_ref}. We use the software package, \textbf{CLASS\_EDE}, of the \textbf{CLASS} code, which includes the modifications for the coupling ($g_{EDE}$) between the EDE field and the CMB photons, to calculate the predicted $EB$ power spectrum.
We then use the MCMC package COBAYA \cite{cobaya:arxiv,cobaya:ads} to extract the best fit value of $g_{EDE}$ for the data. The log-likelihood function used for the fitting is given by, $-2\log(\mathcal{L}) = \sum_b \left[ \mathcal{C}_b^{EB,o} -  \mathcal{C}_b^{EB,p} \right]^2/Var( \mathcal{C}_b)$, where $ \mathcal{C}_\ell^{EB,O}$ and $ \mathcal{C}_\ell^{EB,p}$ are the observed and predicted CMB $EB$ power spectra for the multipole bin, $b$, respectively, and $Var( \mathcal{C}_b)$ is the variance of $\mathcal{C}_b$.

\begin{table}
	\caption{{Best-fit values of the free parameters in the ultra-light axion EDE model with $n$=3 taken from \cite{Poulin:2019}, based on the CMB and BAO data.}}
	\begin{tabular}{|l|c|}
		 \hline $\text { Parameter } $&  Best-fit \\
		\hline $f_{\mathrm{EDE}}$& $ 0.058 $ \\
            $\log _{10}\left(a_c\right) $&$ 3.696 $ \\
            {$\theta_i$}&                      $ 3.0 $\\
            $100 \theta_s $&       $ 1.0414$\\
		$100 \omega_b $&              $ 2.258$\\
		$\omega_{\text {cdm }} $&     $ 0.1299$\\
		$10^9 A_s $&                  $ 2.177$\\
		$n_s $&                       $ 0.988$\\
		$\tau_{\text {reio }}$&       $ 0.068 $\\
		\hline
 		\end{tabular}
\label{tab:bestfit_ref}
\end{table}

\begin{figure}
  \centering
  \includegraphics[width=0.8\textwidth]{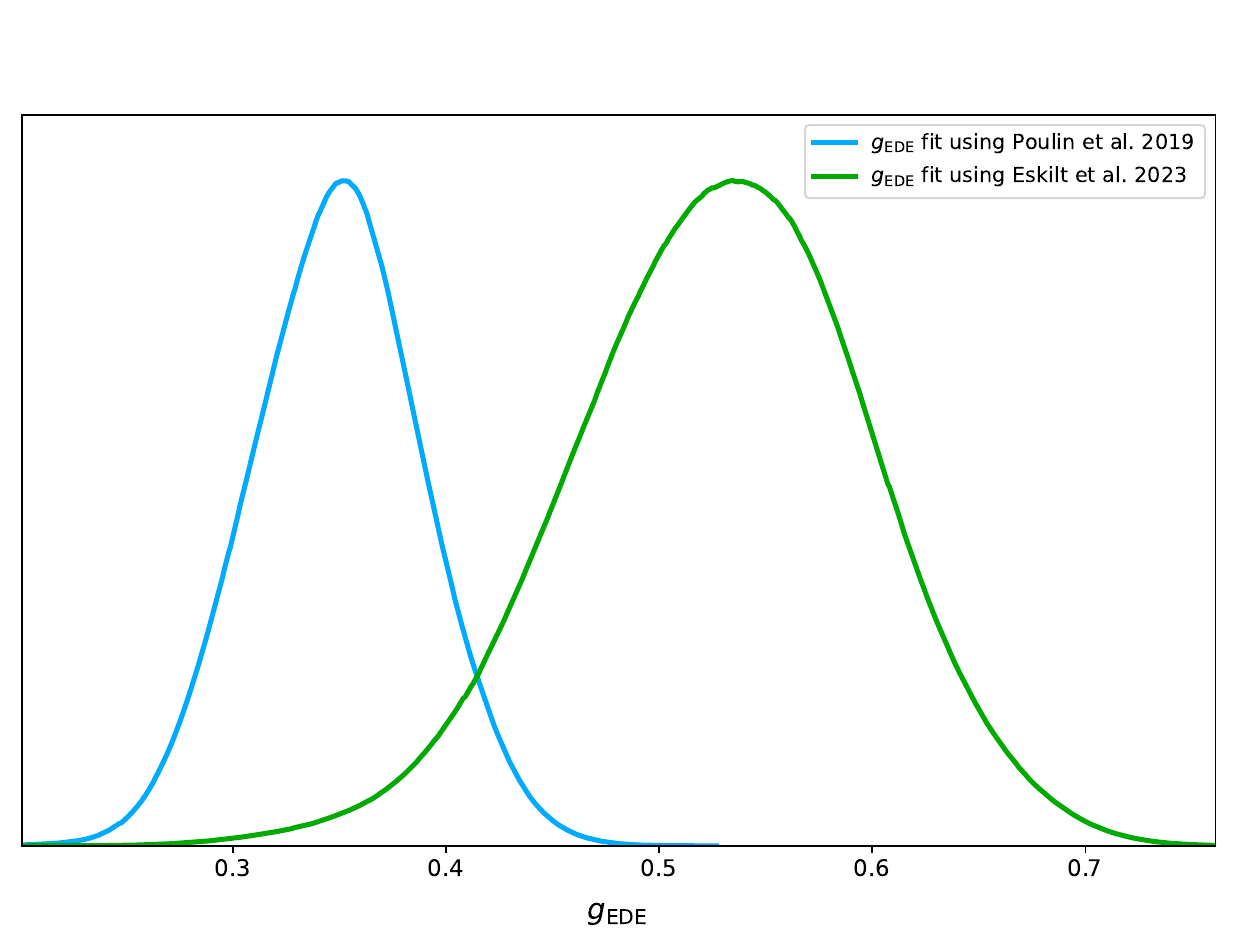}
  \caption{The PDF of $g_{EDE}$ from the CMB $EB$ power spectrum using the best-fit parameter values for the remaining parameters from \cite{Poulin:2019} (blue line), and from \cite{Eskilt:2023} (green line). The corresponding best fit values are $g_{EDE}$=0.3472 and $g_{EDE}$=0.539, respectively.}
  \label{fig:gede_ebonly}
\end{figure}

In Figure \ref{fig:gede_ebonly}, we present the 1-dimensional PDF for $g_{EDE}$ when all other parameters are kept fixed to the values given in Table \ref{tab:bestfit_ref}  (blue line). The best-fit value of $g_{EDE}$ estimated in this way is $0.347$ with $\chi^2 = 64$ for 72 degrees of freedom. 
Additionally, we fit $g_{EDE}$ to the observed CMB $EB$ angular power spectrum, keeping the other parameters fixed to the best-fit values in the second column (Base) of Table I in \cite{Eskilt:2023}. This gives us a best-fit value, $g_{EDE} = 0.549$ with $\chi^2 = 77.5$, and we present the 1-dimensional PDF of $g_{EDE}$ obtained this way in Figure \ref{fig:gede_ebonly}, represented by the green line. We notice that there is a shift in the best-fit value of $g_{EDE}$ when the other cosmological parameters are changed. This result suggests that the fitting of the EDE model to the data may depend on the background cosmological parameters, and studying this dependence is crucial for the fitting process.

\begin{figure}
  \centering
  \includegraphics[width=0.8\textwidth]{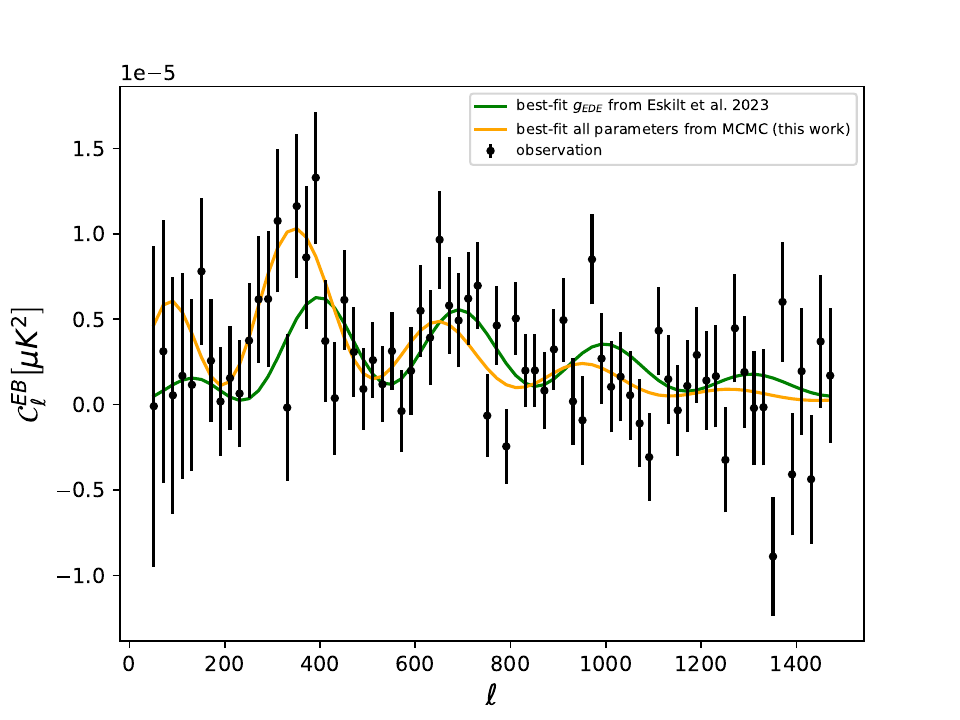}
  \caption{Comparison of the theoretical CMB $EB$ power spectrum by fitting only $g_{EDE}$ keeping all the other parameters fixed to the best-fit results of \cite{Eskilt:2023} (green line), and by fitting all 10 EDE+$\Lambda$CDM parameters with the CMB $TT$, $EE$, $EB$, lensing data, BAO and SH0ES data (orange line). The black points with error bars denote the observed CMB $EB$ power spectrum found in \cite{Eskilt:2022}.}
  \label{fig:ebpower}
\end{figure}

We show a comparison of the predicted CMB $EB$ power spectrum from the EDE model investigated here along with the observations, in Figure \ref{fig:ebpower}. The predicted CMB $EB$ angular power spectrum corresponding to the green line in Figure \ref{fig:gede_ebonly} is denoted by the green line, while the observed power spectrum is represented by the black dots with error bars. The shape of the predicted CMB $EB$ angular power spectrum using the best-fit parameter values from \cite{Eskilt:2023} is in clear disagreement with the observed power spectrum. Our findings from Figure \ref{fig:gede_ebonly} motivate us to look deeper into the dependence of the predicted CMB $EB$ angular power spectrum on the background cosmological parameters, and we discuss our findings in the next section.

\section{Parameter dependance of the CMB $EB$ power spectrum}
\label{sec:eb_param_dep}
\begin{figure}[htb]
\vspace{.5cm}
  \centering
 \includegraphics[width=0.32\textwidth]{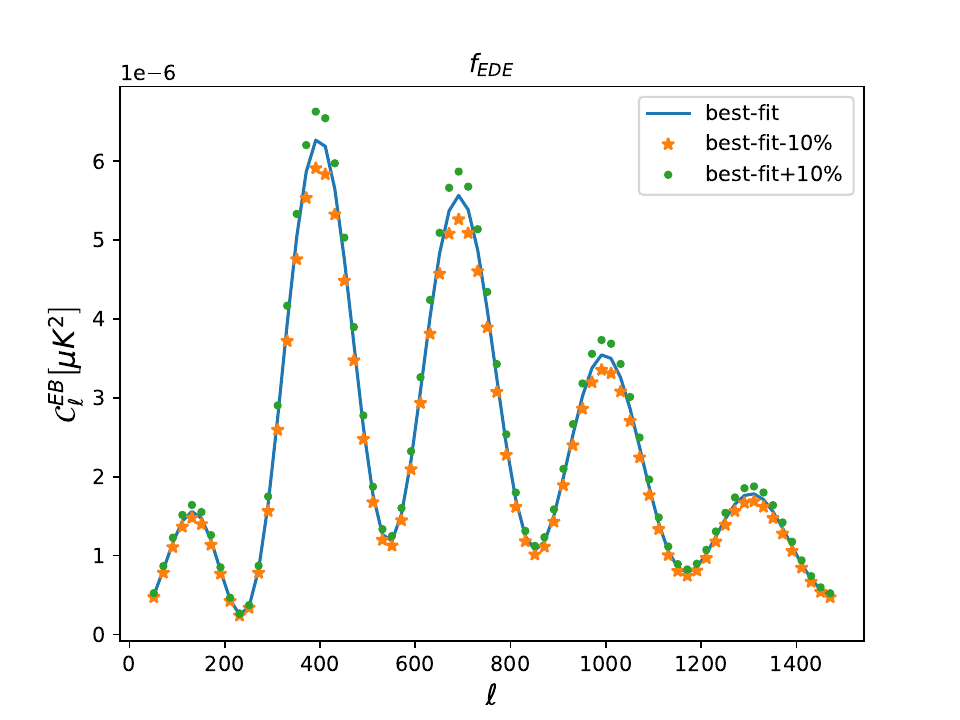} 
 \includegraphics[width=0.32\textwidth]{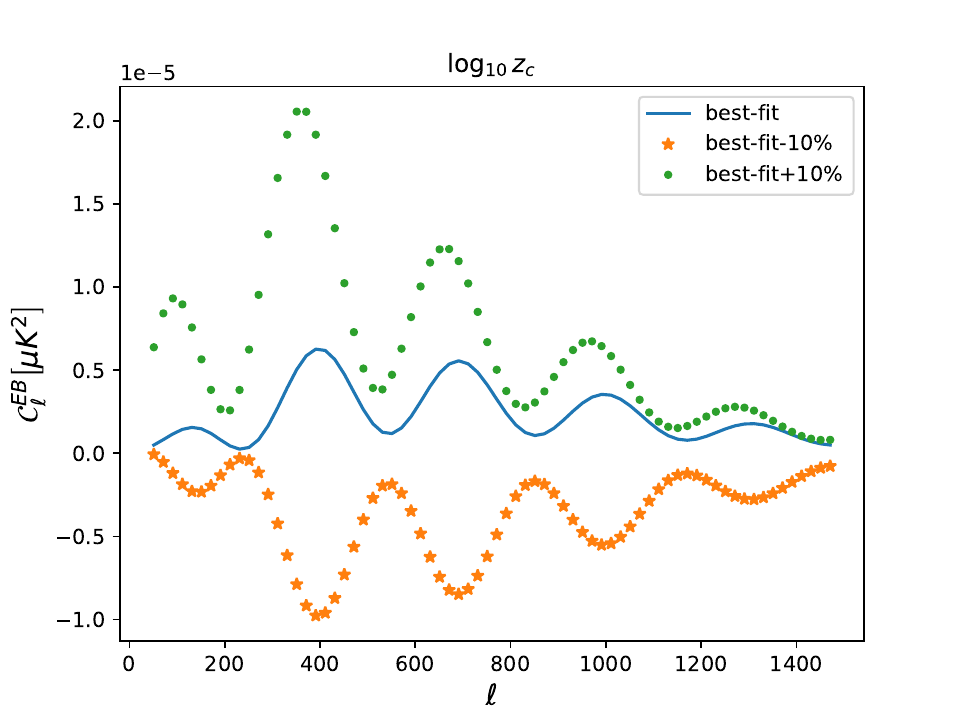}
 \includegraphics[width=0.32\textwidth]{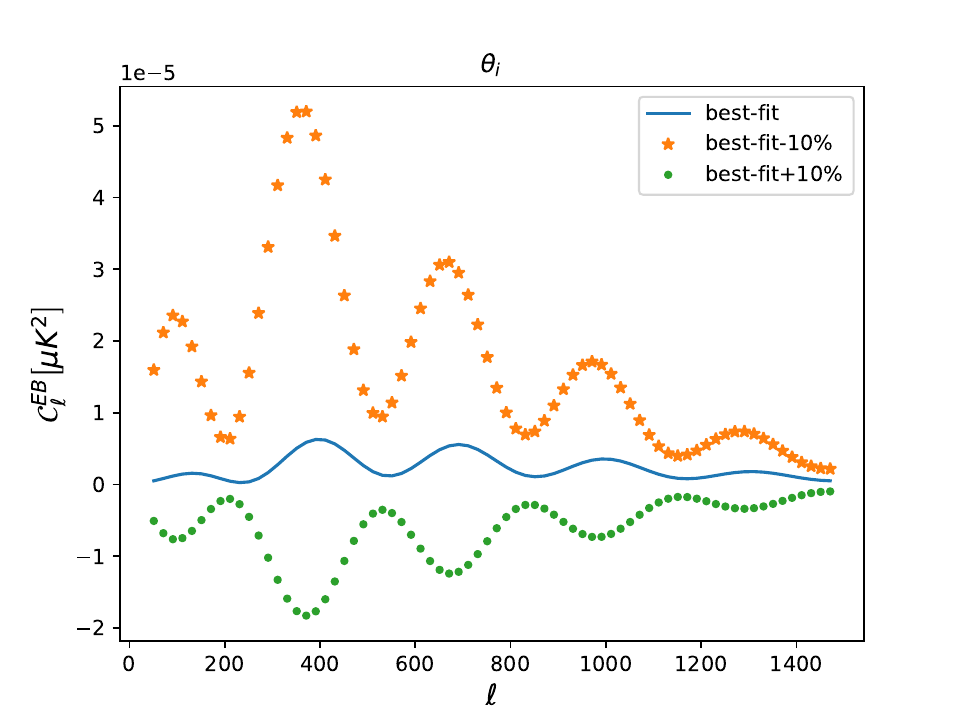} \\
 
 \includegraphics[width=0.32\textwidth]{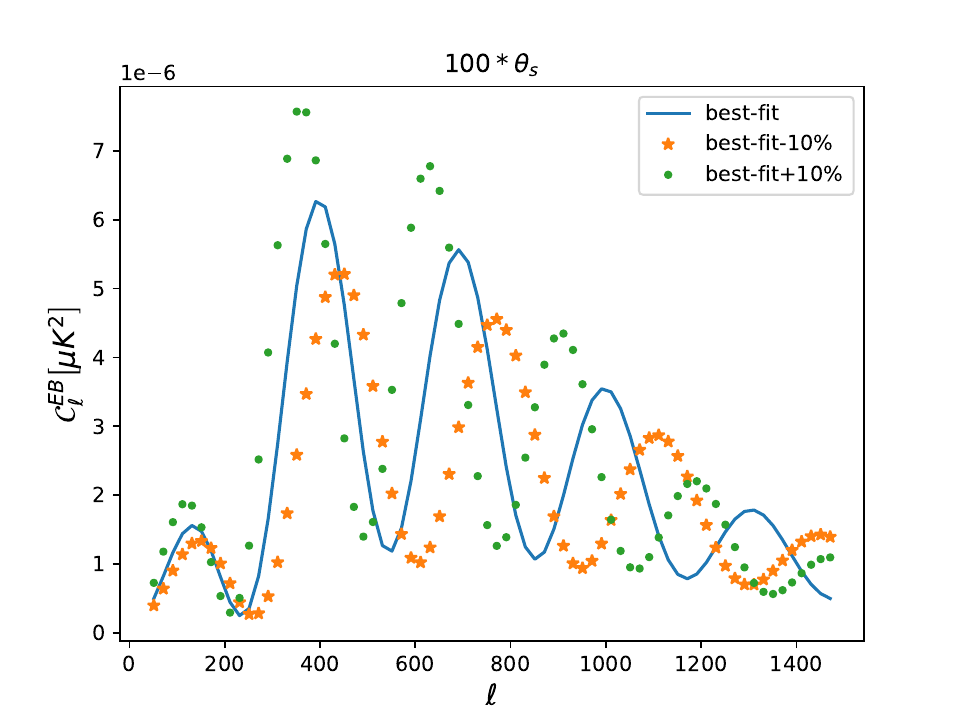} 
 \includegraphics[width=0.32\textwidth]{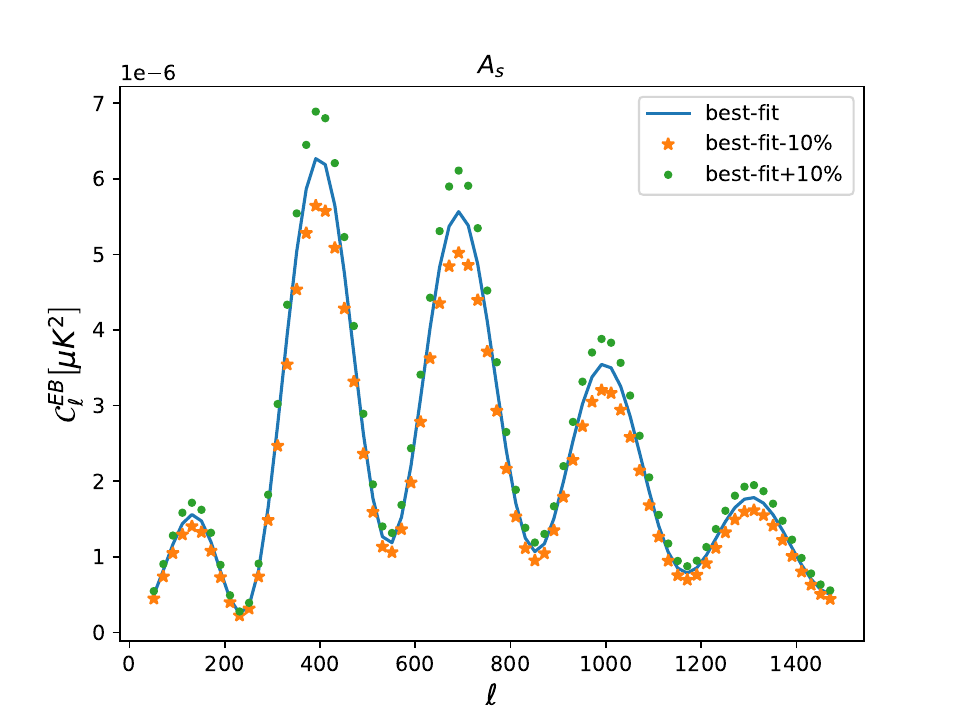} 
 \includegraphics[width=0.32\textwidth]{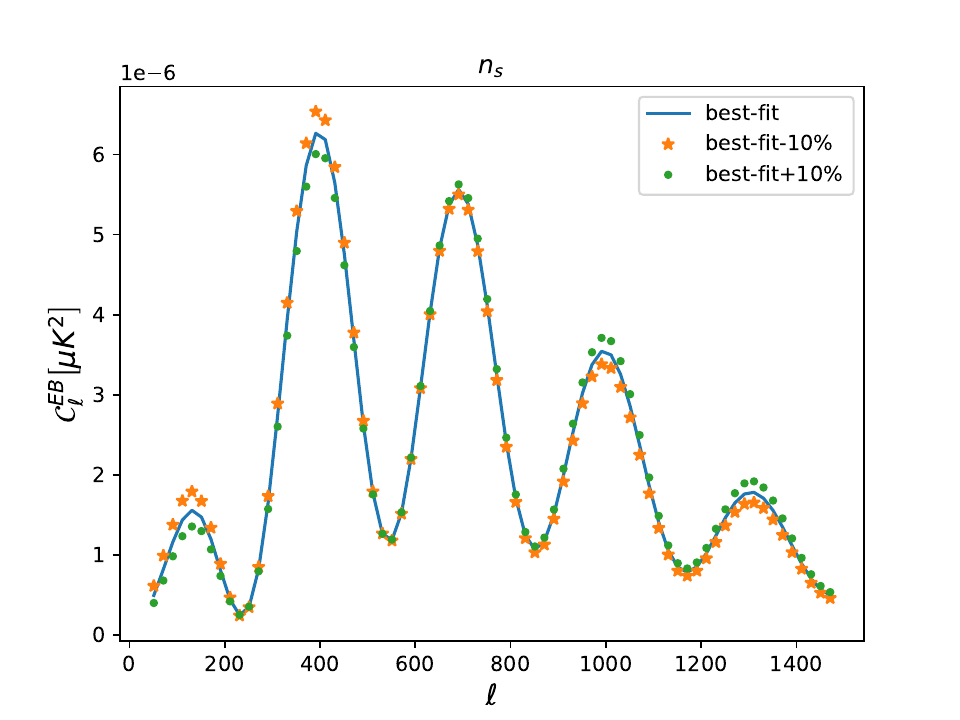}  \\

 \includegraphics[width=0.32\textwidth]{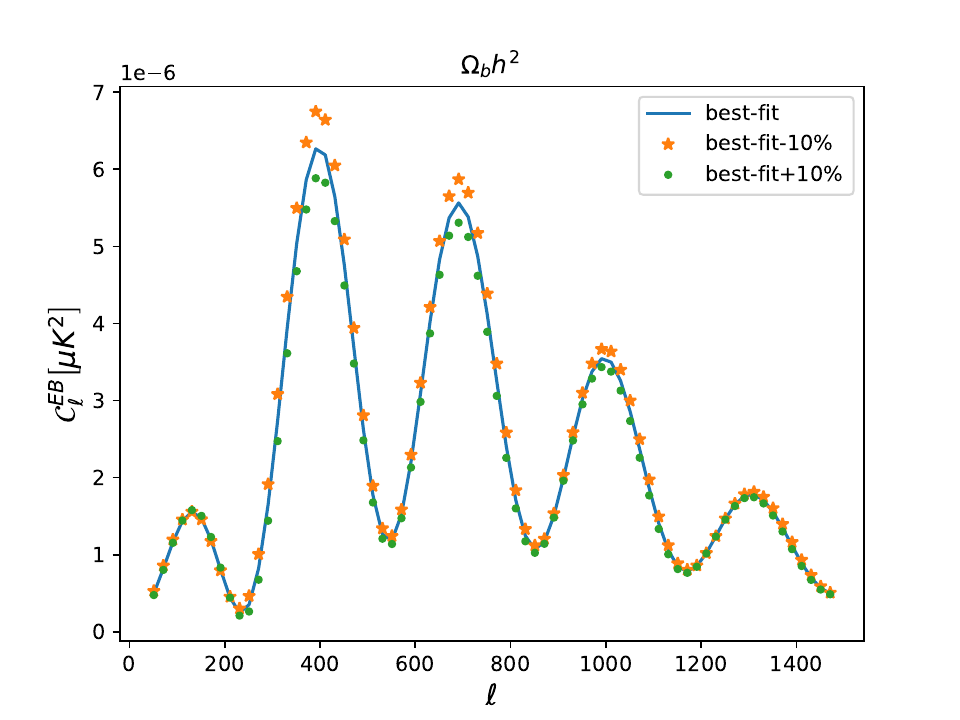} 
 \includegraphics[width=0.32\textwidth]{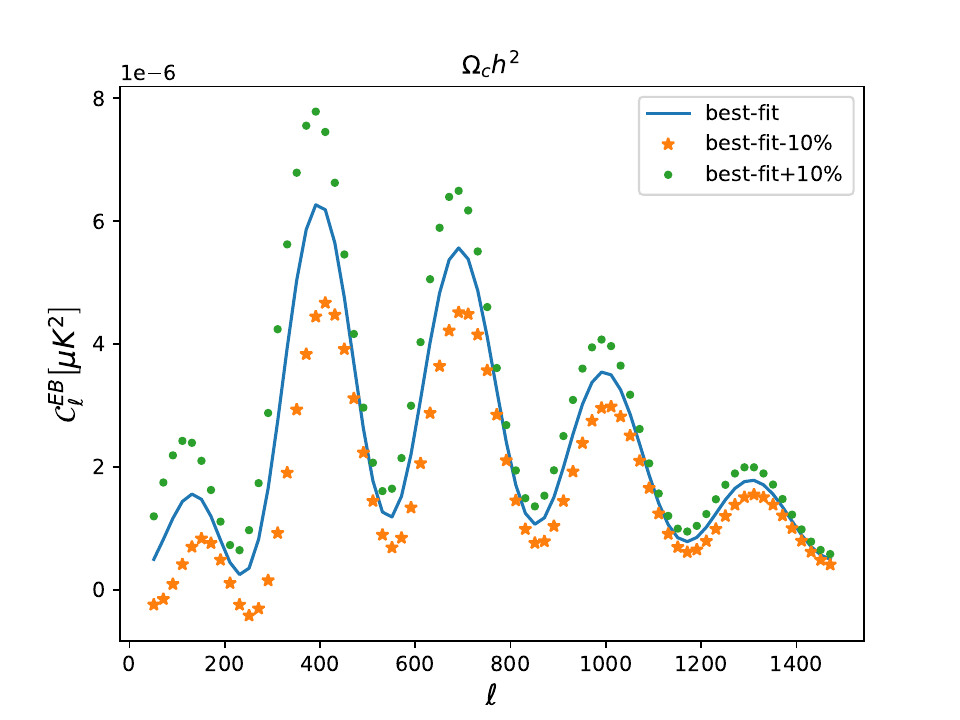} 
 \includegraphics[width=0.32\textwidth]{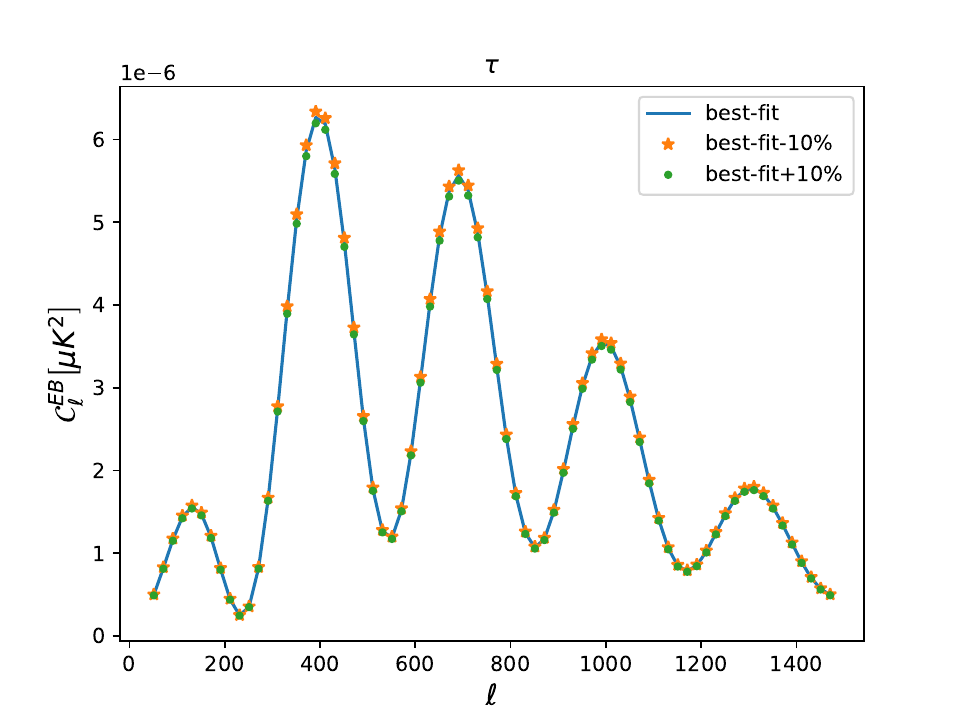}
  \caption{Comparison of the CMB $EB$ power spectra corresponding to a 10\% change in each of the cosmological parameters.}
  \label{fig:eb_param_dep}
\end{figure}

 To test the sensitivity of the shape of the $EB$ spectrum to the cosmological parameters, we run $CLASS\_EDE$ varying one parameter at a time, keeping all other parameters fixed. For each parameter, we use three values, the best-fit value found in the second column (Base) of Table 1 in  \cite{Eskilt:2023}, and $\pm 10\%$ of that value. We fix $g_{EDE}$ to 0.539, which is the corresponding best-fit value for the CMB $EB$ power spectrum. We carry out this process for all nine of the EDE + $\Lambda$CDM parameters except for $g_{EDE}$, and our results are presented in Figure \ref{fig:eb_param_dep}.

In each panel, on the X-axis we have the multipoles $\ell$, and on the Y-axis we have the CMB $EB$ power spectra in $\mu K^2$. The blue lines represent the best-fit value of the corresponding parameter considered in \cite{Eskilt:2023}, the green and orange lines represent the best-fit $+10\%$ and best-fit $-10\%$ values, respectively. From the figure, it is clear that changes to $\log_{10}z_c$, $\theta_i$, $100\theta_s$ and $\omega_{CDM}$ can significantly alter the shape of the CMB $EB$ power spectrum, and the effect of the remaining parameters is largely to change the amplitude of the power spectrum. Considering the lack of precision in the present-day measurements of the CMB $EB$ power spectrum, the bulk of the constraining power lies in the shape of the power spectrum rather than it's amplitude. Therefore, it is important to include $\log_{10}z_c$, $\theta_i$, $100\theta_s$ and $\omega_{CDM}$ in the MCMC exploration of the parameter space when trying to fit the theoretical prediction of the EDE models to the observations. With this understanding, we carry out an MCMC analysis to fit the EDE model to the CMB+BAO+$H_0$ data, in the following section.

\section{Fitting EDE + $\Lambda$CDM model parameters to the CMB data}
\label{sec:fit_ede+lcdm}
In this section, we fit the parameters of the EDE and $\Lambda$CDM models to the CMB and lensing data provided by Planck \cite{planck2018:spectra, planck2018:lensing}, BAO data from SDSS DR12 \cite{bao:2017}, and local measurements of the Hubble constant \cite{RiessH0:2020}. We vary four EDE parameters and six $\Lambda$CDM parameters for a total of ten parameters. The ten parameters in our case are: $f_{EDE}$, $g_{EDE}$, $\log_{10}z_c$, $\theta_i$, $100*\theta_s$, $A_s$, $n_s$, $\omega_b$, $\omega_{CDM}$ and $\tau$. We use flat priors for all the parameters while exploring the parameter space with MCMC software package Cobaya \cite{mcmc:2002, mcmc:2013}. We choose the limit $R-1<0.05$ for the Gelman-Rubin convergence criterion \cite{Gelman:1992}, and consider the chains converged when this condition is satisfied.

\begin{table}
	\caption{Best-fitting (with 68\% limits) and marginalised values of the 10 parameters in the ultra-light axion EDE model with $n$=3 \cite{Murai:2023}, using CMB $TT$, $EE$, $EB$, lensing data, BAO data from SDSS DR12, and $H_0$ measurements from the SH0ES team.}
	\begin{tabular}{|l|c|c|}
		 \hline $\text { Parameter } $&  Best-fit & Marginalised \\
		\hline & & \\
            $f_{\mathrm{EDE}}$ & $ 0.1950^{+0.0579}_{-0.0774} $ & $0.1801$\\
             & & \\
            $g_{\mathrm{EDE}}$ & $0.1483^{+0.2942}_{-0.1021}$ & $0.1701$ \\
             & & \\
            $\log _{10}\left(z_c\right) $&$ 3.4752^{+0.0853}_{-0.0606} $ & $3.4869$ \\
             & & \\
            {$\theta_i$}&                      $ 1.89^{+0.8703}_{-0.5302}$ & $2.0733$ \\
             & & \\
            $100 \theta_s $&       $ 1.0404^{+0.0014}_{-0.0017}$ & $1.0405$ \\
             & & \\
		$100 \omega_b $&              $ 2.272^{+0.09}_{-0.102}$ & $2.259$ \\
             & & \\
		$\omega_{\text {cdm }} $&     $ 0.1558^{+0.0131}_{0.0157}$ & $0.1523$ \\
             & & \\
		$10^9 A_s $&                  $ 2.3414^{+0.0933}_{-0.068}$ & $2.339$ \\
             & & \\
		$n_s $&                       $ 0.9887^{+0.0213}_{-0.0213}$ & $0.9845$ \\
             & & \\
		$\tau_{\text {reio }}$&       $ 0.0679^{+0.0139}_{-0.0097}$ & $0.0711$ \\
             & & \\
            $H_0$ & $72.03^{+2.73}_{-3.41}$ & $71.26$ \\
             & & \\
		\hline
 		\end{tabular}
\label{tab:bestfit}
\end{table}

\begin{figure}
  \centering
  \includegraphics[width=1.0\textwidth]{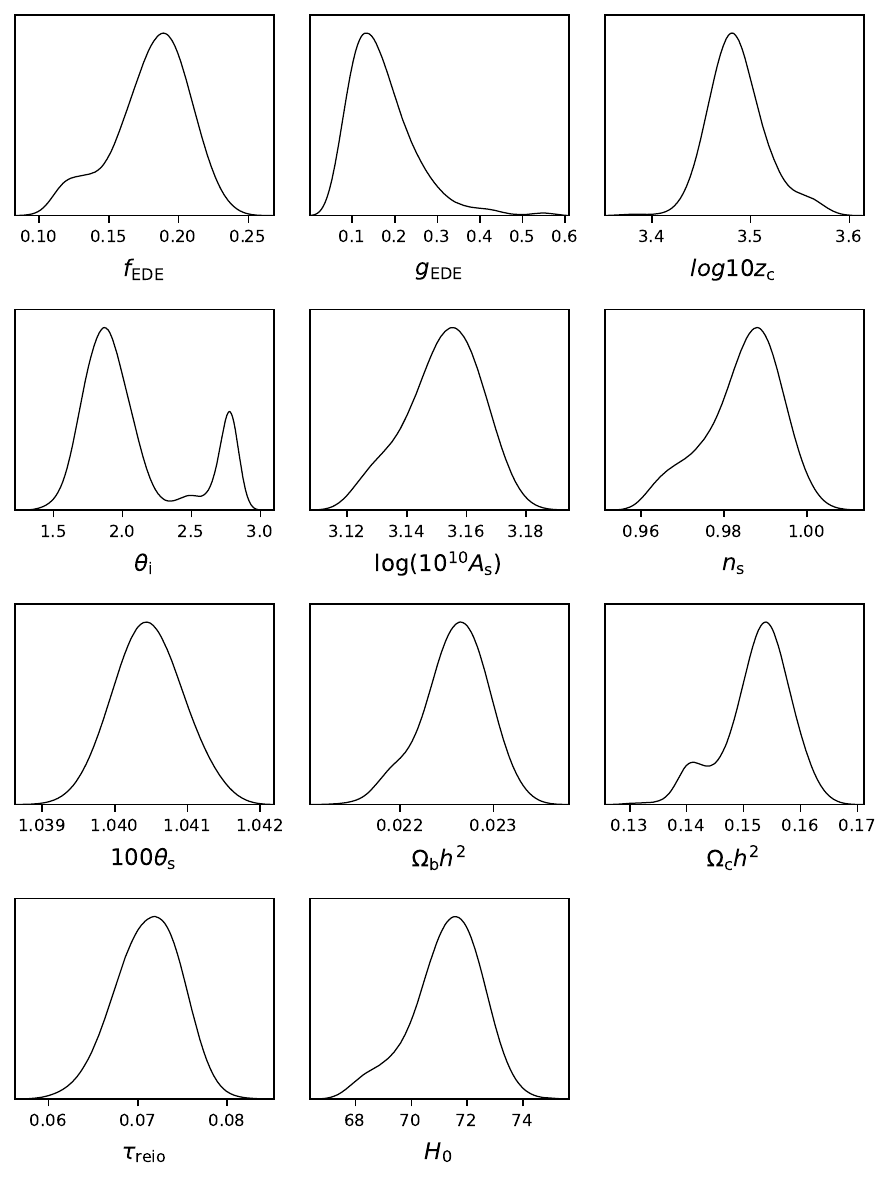}
  \caption{The 1D marginalised PDFs of the 10 fitted EDE+$\Lambda$CDM parameters, along with $H_0$ (derived), from the MCMC fitting of CMB, BAO and SNIa.}
  \label{fig:1d_pdfs}
\end{figure}

\begin{figure}
  \centering
  \includegraphics[width=1.0\textwidth]{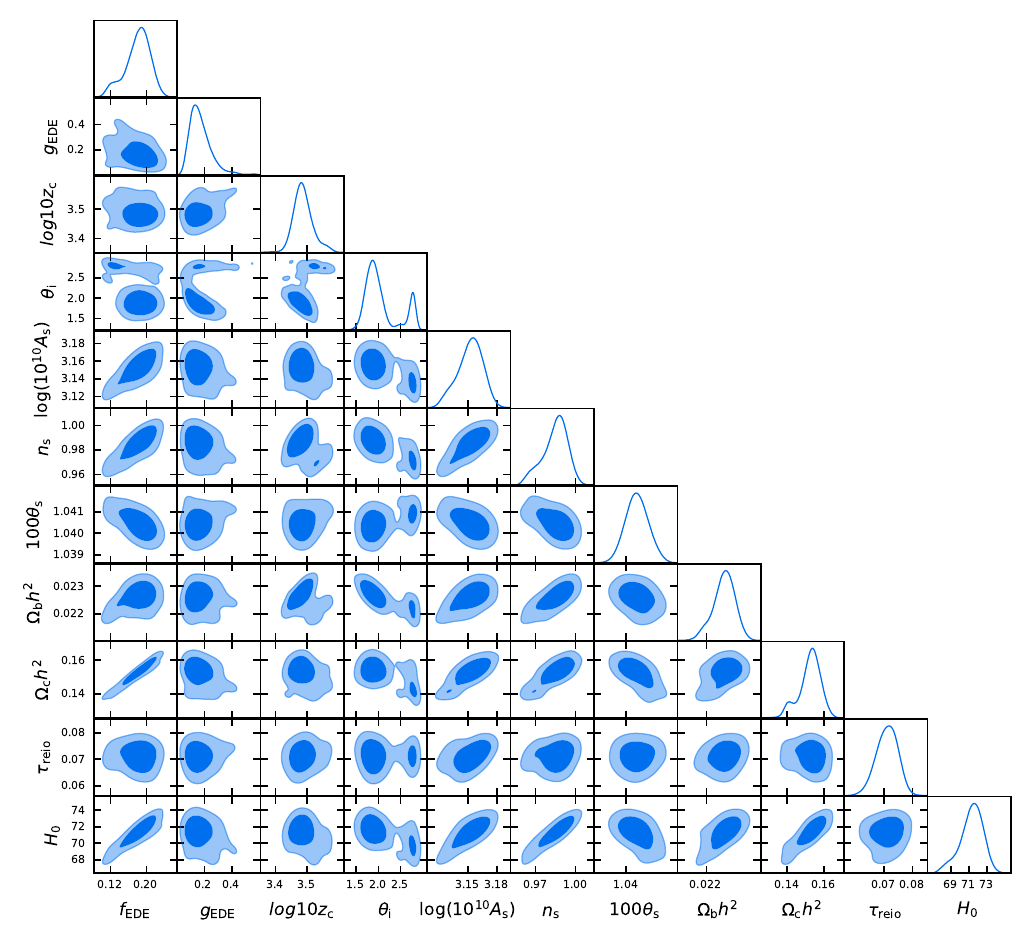}
  \caption{The 2D posteriors of the 10 fitted EDE+$\Lambda$CDM parameters, along with $H_0$ (derived), from the MCMC fitting of CMB, BAO and SNIa.}
  \label{fig:2d_pdfs}
\end{figure}

We present our results for the 1-D and 2-D marginalised posterior distributions from the parameter estimation of EDE+$\Lambda$CDM parameters in Fig. \ref{fig:1d_pdfs} and Fig. \ref{fig:2d_pdfs}, respectively. We write the best-fit parameter values and marginalised means along with the corresponding 68\% limits in Table \ref{tab:bestfit}. We find a relatively large value of $f_{EDE}$ than some other works in the literature, and a consequently larger value of $H_0$. Our best-fit $H_0$ is consistent with the SH0ES result \cite{Riess:2022} within the 68\% limits, showing that axion-like Early Dark Energy is a viable candidate for explaining the Hubble tension. In the 2-D posteriors, we find that the $\theta_i$ panels show multimodal distributions, but all the other panels have well-behaved posteriors. 

Next, we compare the predictions of the EDE+$\Lambda$CDM model with our best-fit parameters, with the observed CMB $EB$ power spectrum in Fig. \ref{fig:ebpower}. We find that contrary to the results in \cite{Eskilt:2023}, the predicted CMB $EB$ power spectrum from axion-like EDE is in good agreement with the observations, with a $\chi^2$ of 68 for 72 degrees of freedom. This difference comes from allowing the background cosmological parameters to vary, where \cite{Eskilt:2023} kept their values fixed. As shown in section \ref{sec:eb_param_dep}, changes in the background parameter values can significantly change the shape of the theoretical $EB$ power spectrum.

\section{Discussion}
\label{sec:7}
We have revisited the axion-like Early Dark Energy model with $n$=3, in the context of the Hubble tension and the CMB $EB$ power spectrum, which is the signal of cosmic birefringence in the CMB data. The coupling between the EDE field and the CMB photons give rise to a non-vanishing $EB$ power spectrum. While recent results suggest that the axion-like EDE model with $n$=3 is not {favored} as the explanation for the origin of cosmic birefringence,  we have shown that this is not necessarily true, and depends on one's choice of cosmological parameters. We explore the full 10-dimensional parameter space of the $n$=3 $EDE$ +$\Lambda$CDM model with the publicly available MCMC software COBAYA \cite{cobaya:arxiv, cobaya:ads}. We use CMB temperature data, $E$ mode polarisation data, $EB$ cross-correlation data and lensing data from Planck, BAO data from SDSS dr12, and $H_0$ measurements from the SH0ES team. We obtain best-fit values for model parameters that are in agreement with the observations of cosmic birefringence, and are also consistent with the late Universe measurements of the Hubble constant. Our results show that the $n$=3 axion-like EDE model can simultaneously explain the observation of cosmic birefringence, as well as resolve the Hubble tension. In our analysis for the $n=3$ model, the best-fit value of the rotation angle $\beta$ from Equation \ref{eqn:beta} is $0.059^{\circ}$.

The large error bars in the measurement of the CMB $EB$ power spectrum make it difficult to claim that the axion-like EDE is the correct explanation for the observed cosmic birefringence. Observations from LiteBIRD are expected to improve the significance of the measurement of cosmic birefringence from 3.6-$\sigma$ to the level of detection \cite{Litebird:2022,Minami:2020b}. The data from the AliCPT-1 mission will reduce the uncertainties of the observed CMB $EB$ power spectrum by $\approx$ 25\% \cite{Dou:2024}, but these will not greatly enhance the constraints on EDE models. The availability of more precise data in the future will have the final word in this regard, but for now EDE remains a viable candidate for providing a unified explanation for the Hubble tension and cosmic birefringence. If a redshift depedence of the rotation angle $\beta$ is observed in agreement with equation \ref{eqn:beta}, then that could be evidence in {favor} of string theory.

\backmatter

\bmhead{Acknowledgements}

We sincerely thank Stephen Appleby for the very insightful discussions on the topic, and for for providing support for the numerical calculations. We also thank Patricia Diego-Palazuelo and Toshiya Namikawa for their suggestions for improving the manuscript. Some of the simulations in this article were carried out on the Shakti cluster at Manipal Centre for Natural Sciences (MCNS). L.Yin was supported by an appointment to the YST Program at the APCTP through the Science and Technology Promotion Fund and Lottery Fund of the Korean Government. J. K. was supported by the MCNS faculty development fund. J. K. was also supported by an appointment to the Junior Research Group Program at the APCTP through the Science and Technology Promotion Fund and Lottery Fund of the Korean Government, and by the Korean Local Governments in Gyeongsangbuk-do Province and Pohang City. The  work of B.-H. Lee is partially supported by the National Research Foundation of Korea (NRF) grant 2020R1F1A1075472.

\end{document}